# mTORC1 regulates cytokinesis through activation of Rho-ROCK signaling

Timothy R. Peterson[1], Mathieu Laplante[1], Van Veen E[2], Van Vugt MA[2], Thoreen CC[1], David M. Sabatini[1].

[1]Whitehead Institute for Biomedical Research, Howard Hughes Medical Institute, and Massachusetts Institute of Technology, Department of Biology, Nine Cambridge Center, Cambridge, MA 02142.
[2]MIT Center for Cancer Research, 77 Massachusetts Avenue, Cambridge, MA 02139.

## Summary

Understanding the mechanisms by which cells coordinate their size with their ability to divide has long attracted the interest of biologists. The Target of Rapamycin (TOR) pathway is becoming increasingly recognized as a master regulator of cell size, however less is known how TOR activity might be coupled with the cell cycle. Here, we establish that mTOR complex 1 (mTORC1) promotes cytokinesis through activation of a Rho GTPase-Rho Kinase (ROCK) signaling cascade. Hyperactivation of mTORC1 signaling by depletion of any of its negative regulators: TSC1, TSC2, PTEN, or DEPTOR, induces polyploidy in a rapamycin-sensitive manner. mTORC1 hyperactivation-mediated polyploidization occurs by a prolonged, but ultimately failed attempt at abcission followed by re-fusion. Similar to the effects of ROCK2 overexpression, these mTORC1-driven aberrant cytokinesis events are accompanied by increased Rho-GTP loading, extensive plasma membrane blebbing, and increased actin-myosin contractility, all of which can be rescued by either mTORC1 or ROCK inhibition. These results provide evidence for the existence of a novel mTORC1-Rho-ROCK pathway during cytokinesis and suggest that mTORC1 might play a critical role in setting the size at which a mammalian cell divides.

## Introduction

Key to the proper transition of cells through the cell cycle is their passage through multiple signaling 'checkpoints'. One of the first insights into the existence and function of one of these checkpoints was the isolation and characterization of a mutation in the Wee1 gene. Wee1 mutant cells were shown to abrogate a previously unknown to exist cell size checkpoint strikingly, undergoing cell division at half the size of wild-type cells (*1*). Since this discovery, our understanding of the mechanisms controlling cell size and cell division have rapidly, though separately, evolved. The Rho pathway has emerged as a central player in controlling the actin-myosin-based contractile events required for cytokinesis (*2, 3*). Regarding the regulation of cell size, the TOR signaling pathway has received considerable attention, in particular for its role in controlling the G1 growth phase of the cell cycle. Though it has largely become known as a controller of protein synthesis and the G1 program, it has also long been clear that TOR has a second, distinct essential function involving the cell-cycle dependent organization of the cytoskeleton (*4*). Mutations in TOR2 do not arrest cells in G1 as is the case with rapamycin treated cells, but rather cause arrest at multiple points in the cell cycle (*5*) accompanied by profound actin cytoskeletal disorganization (*4*). That the lethality of a TOR2 mutation could be rescued by overexpression of components of the Rho pathway hinted at the relationship between these pathways during the cell cycle (*6*). However, the importance of such a coupling during cell division still remains largely unknown. To address this question, we have investigated the involvement of mTOR signaling in cytokinesis. We find that mTORC1 hyperactivation causes prolonged cytokinesis due to Rho-ROCK hyperactivity and is associated with increased polyploidy, a condition thought to underlie the malignancy of certain cancers.

## Results

The mTOR pathway is repressed through the actions of multiple upstream inhibitory proteins including TSC1, TSC2, PRAS40, PTEN, and REDD1 (*7*). We recently identified a novel negative regulator of the mTOR pathway, DEPTOR, which binds mTOR and inhibits both mTORC1 and mTORC2 signaling pathways (*8*). In characterizing DEPTOR function, we noticed that its depletion in cells by RNAi promoted their multinucleation (Fig. 1A). To quantify this phenomenon, we measured the DNA content of DEPTOR knockdown cells compared with control cells. The results were clear; DEPTOR knock-down increased the percentage of 8N cells relative to those of control knockdown cells (Fig. 1B). Because DEPTOR depletion activates both the mTORC1 and mTORC2 pathways, we sought to determine whether either of these pathways might be responsible for mediating the effects of DEPTOR on ploidy. First, we knocked down known regulators of mTORC1 and/or mTORC2: raptor, rictor, mTOR, TSC1, TSC2, and PTEN and measured their DNA content. Similar to that of DEPTOR knockdown cells, cells with knockdown of any of the negative regulators of mTOR: TSC1, TSC2, or PTEN increased the percentage of 8N cells relative to control knockdown cells (Fig. 1B). Fittingly, the increased >4N DNA content of TSC1 knockdown was similar to those seen in a different cell type with complete loss of TSC1 (*9*). That neither mTORC1 nor mTORC2 inactivation by knockdown or raptor or rictor, respectively increased the percentage of 8N cells is consistent with DEPTOR being a negative regulator of mTOR signaling (Fig. 1B). Finally, treatment of DEPTOR or TSC2 knockdown cells with rapamycin, which only significantly inhibits mTORC1 and not mTORC2 in the cells we assessed (*8*), fully reduced the percentage of 8N cells (Fig. 1B) suggesting that deregulation of mTORC1 is responsible for effects of mTOR on polyploidy.

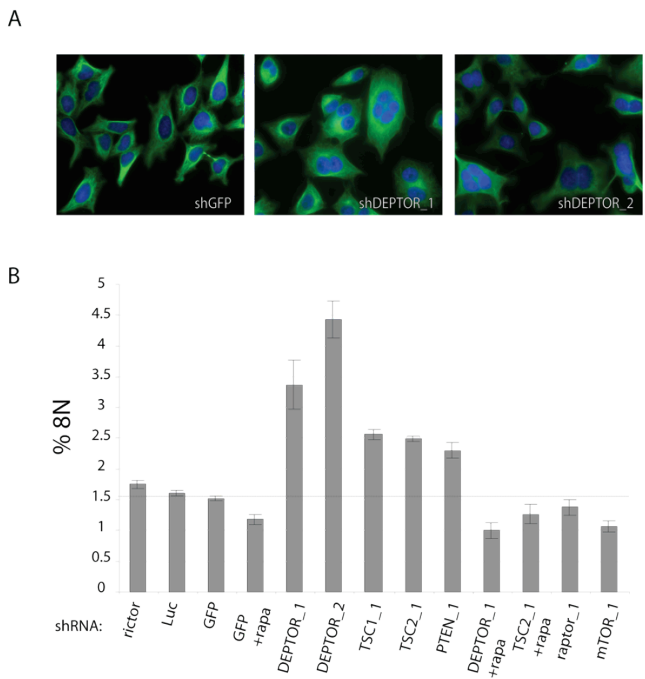

**Fig. 1. mTORC1 hyperactivation causes polyploidy.** (A) HeLa cells were infected with lentivirus expressing shRNAs targeting DEPTOR or GFP. Cells were processed in an immunofluorescence assay to detect α-tubulin (green), costained with DAPI for DNA content (blue), and imaged. (B) HeLa cells pretreated with 100nM rapamycin or vehicle were infected with lentivirus expressing the indicated shRNAs, stained with propidium iodide, and analyzed for DNA content by flow cytometry. ≥ 20,000 cells were analyzed for each condition. Error bars indicate standard error for n=3.

To explore at which point in their cell cycle mTORC1-hyperactivated cells became polyploid, we visually monitored individual knockdown cells over time. Both wild-type and DEPTOR knockdown cells progressed through interphase and through early stages of mitosis similarly (Fig. 2A). However, upon daughter cell separation, the surface of DEPTOR knockdown cells, unlike control cells, began to bleb massively and this was subsequently followed after a prolonged period of blebbing by fusion of their membranes and the creation of one cell containing both daughter nuclei (Fig. 2A). The same prolonged period of surface blebbing pattern was also observed in cells deficient in TSC1, TSC2, or PTEN but not in rictor knockdown cells (Fig. 2B-C) suggesting that blebbing-associated multinucleation is due to hyperactivation of mTORC1 and not mTORC2 (either directly or indirectly through the well-characterized mTORC1-PI3K feedback loop(*10*)). To strengthen the claim that this aberrant surface blebbing was mediated by mTORC1, we treated DEPTOR knock-down cells with rapamycin immediately preceding the point at which they began to separate into two daughter cells. Rapamycin treatment dramatically reduced the time required for the separation of DEPTOR-deficient daughter cells as well as the time of their surface blebbing (Fig. 2B, 2D). Correlating with the degree of polyploidy, these results suggest that surface blebbing-associated cytokinesis defects in DEPTOR deficient cells are the result of mTORC1 hyperactivation. During cytokinesis, cytoskeletal rearrangements are critical for physically separating cells after their duplication (*11*). Both microtubules and the actin are core components of the cytoskeleton driving this process (*11*), therefore we assessed their morphology with or without mTORC1 hyperactivation during multiple stages of cell cycle. Control and DEPTOR knockdown cells had similar cortical actin positioning and assembly of their α-tubulin-containing spindles in mitosis (Fig. 2E). However, in cytokinesis, while control daughter cells had symmetrical actin cortices, in DEPTOR depleted cells, their actin architecture was grossly disorganized and appeared to be forming blebbing patterns similar to that which we observed in DEPTOR depleted cells by light microscopy (Fig.2A-B, 2E).

Actin-myosin contractility provides critical mechanical force to promote cell cleavage in cytokinesis and is known to be coordinated by many Rho-GTP-dependent kinases which regulate this contractility by phosphorylating myosin light chain 2 (MLC2) (*12*). To test a role for mTORC1 in regulating actin-myosin contractility, we first confirmed in our cell system that the MLC2 phosphorylating kinase, Rho kinase 2 (ROCK2), regulates the morphology of the actin cytoskeleton (*13*). While overexpression of kinase dead ROCK2 did not strongly perturb the actin cytoskeleton compared with overexpression of a control protein, overexpression of wild-type ROCK2 produced strong actin blebbing much like that we had seen with DEPTOR knockdown cells (Fig. 2E). As expected, ROCK2 overexpression also increased S19 MLC2 phosphorylation (Fig. 3A). To test whether DEPTOR regulated ROCK signaling during cytokinesis, we measured MLC2 phosphorylation in DEPTOR and control knockdown cells. Typical of metazoan cells in cytokinesis, we detected prominent S19 MLC2 phosphorylation in control cells that was reduced by inhibition of ROCK activity (Fig. 3B) (*14, 15*). On the other hand, DEPTOR depleted cells, had substantially elevated levels of S19 MLC2 phosphorylation compared with control cells (Fig. 3C). That the effects of DEPTOR loss of S19 MLC2 phosphorylation were strongly reduced by ROCK inhibition (Fig. 3B) suggests that DEPTOR loss activates ROCK signaling.

Because it is possible that DEPTOR knockdown activates ROCK signaling independent of its effects on

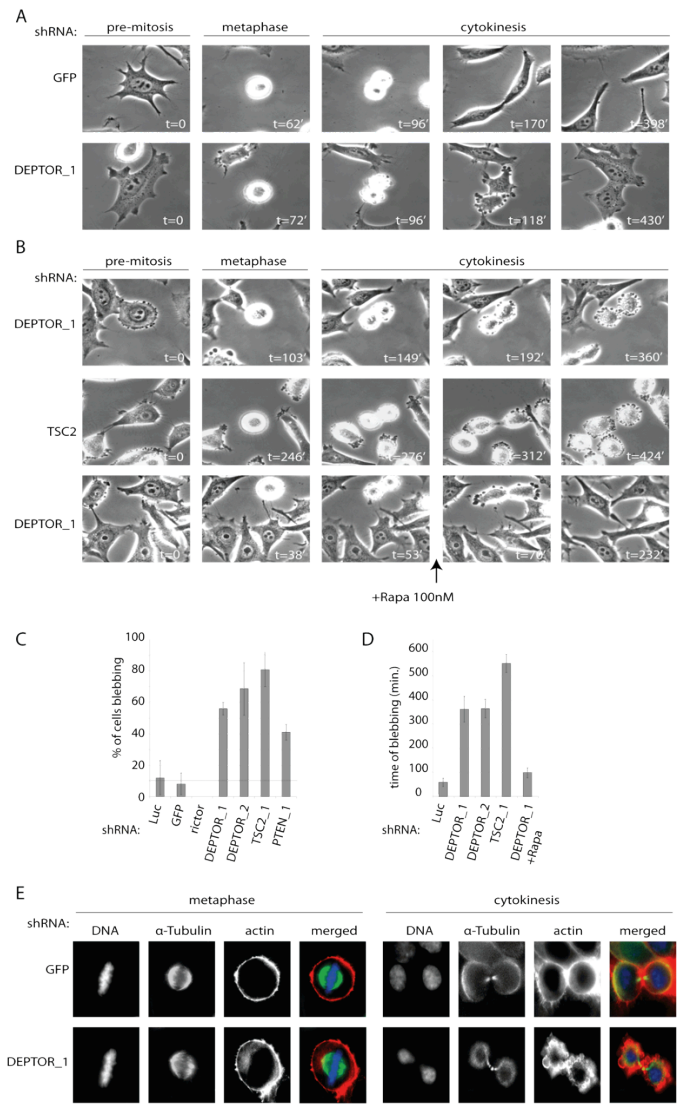

**Fig. 2. mTORC1 hyperactivation distorts the actin cytoskeleton and prolongs cytokinesis**.
(A) and (B) HeLa cells were infected with lentivirus expressing shRNAs targeting DEPTOR, TSC2, or GFP were analyzed by light microscopy (10x). Images were captured once per minute. (C) and (D) All images from (A) and (B) were visually inspected and manually counted to determine the % of cells blebbing and time of blebbing in minutes. ≥ 50 cells were analyzed for each sample. Error bars indicated standard error for n=3. (E) HeLa cells were infected with lentivirus expressing a shRNA targeting DEPTOR or GFP. Cells were then processed in an immunofluorescence assay to detect α-tubulin (green), actin (red), costained with DAPI for DNA content (blue), and imaged.

mTORC1, we tested whether mTORC1 inhibition would reverse the effects of DEPTOR depletion on MLC2 phosphorylation. Indeed, treating DEPTOR knockdown cells with either rapamycin or a concentration of Torin1 which only potently inhibits mTORC1 signaling (*8*) reduced S19 MLC2 phosphorylation (Fig. 3C). We next sought to determine whether the effects of DEPTOR on ROCK activity were direct. Because ROCK is activated by binding to the active, GTP-bound form of Rho (*16*) and because TOR is known to regulate Rho GEF activity (*6*), we tested whether DEPTOR knockdown regulates ROCK by altering the Rho-GTP/GDP state. DEPTOR knockdown cells had higher levels of Rho-GTP than control cells and this increased activity was reduced by rapamycin (Fig. 3D). These results are consistent with mTORC1 activating ROCK upstream of ROCK itself by promoting Rho activity. Because ROCK directly phosphorylates many substrates besides MLC2, we checked whether other ROCK outputs were also regulated by manipulation of mTORC1. Mypt1 is the targeting subunit for the S19 MLC2 phosphatase and is known to be inhibited in its activity by phosphorylation at its T696 site by ROCK (*17, 18*). DEPTOR knockdown increased and mTOR knockdown decreased Mypt1 T696 phosphorylation consistent with mTORC1 positively regulating ROCK activity toward this substrate (Fig. 3E). Another target of ROCK pathway is Cofilin. Cofilin is an F-actin severing protein which is inhibited through its S3 phosphorylation by the ROCK-activated kinase, LIMK1 (*19*). Consistent with the effects of DEPTOR knockdown on MLC2 and Mypt1 phosphorylation, S3 cofilin phosphorylation was increased by DEPTOR knockdown above the levels seen in control knockdown in multiple cell types (Fig. 3F). In total, these results suggest that Rho-ROCK signaling toward multiple effector pathways is positively regulated by mTORC1.

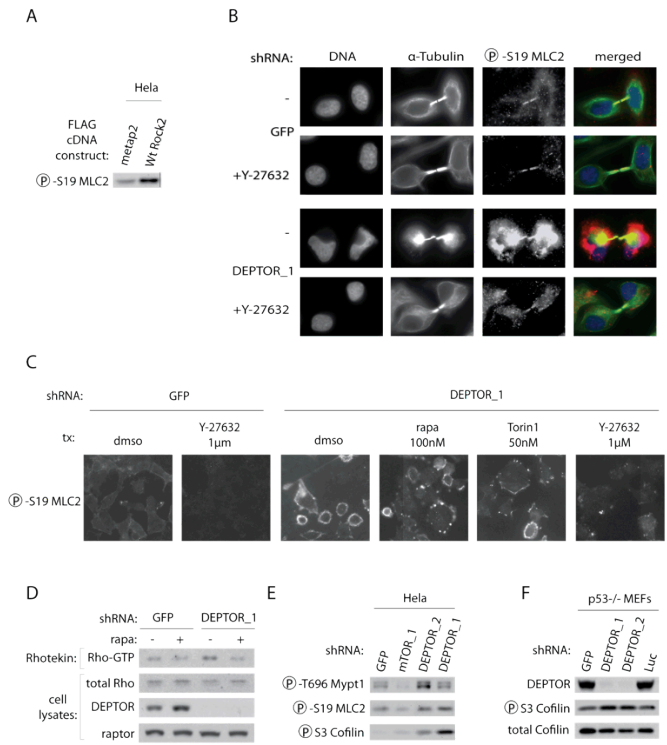

**Fig. 3. mTORC1 hyperactivation activates Rho-ROCK signaling.**
(A) HEK-293T cells transfected with the indicated cDNAs were analyzed by immunoblotting for S19 MLC2 phosphorylation. (B) HeLa cells were infected with lentivirus expressing shRNAs targeting DEPTOR or GFP, treated for 10 minutes with 10µM of Y-27632 or vehicle, and processed in an immunofluorescence assay to detect α-tubulin (green), S19 MLC2 phosphorylation (red), costained with DAPI for DNA content (blue), and imaged. (C) HeLa cells were infected with lentivirus expressing shRNAs targeting DEPTOR or GFP, treated for 10 minutes with 100nM rapamycin, 50µM Torin1, 10µM of Y-27632, or vehicle and processed in an immunofluorescence assay to detect S19 MLC2 phosphorylation and imaged. (D) HeLa cells were infected with lentivirus expressing shRNAs targeting DEPTOR or GFP, treated for 10 minutes with 100nM rapamycin or vehicle, immunoprecipitation of Rhotekin binding proteins was performed, and immunoprecipitates and lysates were analyzed by immunoblotting for the indicated levels of the specified proteins. (E) HeLa cells were infected with lentivirus expressing the indicated shRNAs were analyzed by immunoblotting for the phosphorylation states of the specified proteins. (F) Indicated p53-/- MEFs were generated and analyzed as in (E).

To better define how DEPTOR-mTORC1 controls Rho-ROCK activity, we assessed DEPTOR localization in cells undergoing cytokinesis. We detected DEPTOR throughout the cell, though it was clearly enriched at the centrosomes and the cleavage furrow, two subcellular localizations, which interestingly are also enriched for ROCK and RhoA, respectively (Fig. 4A) (*20, 21*). Previously, we have shown that DEPTOR phosphorylation state and levels are regulated by growth factors in an mTOR-dependent manner (*8*), therefore we assessed whether DEPTOR was also regulated in a cell cycle-dependent manner. Arresting cells in mitosis with nocodazole, a microtubule disrupting agent (*22*), significantly impaired the gel mobility of the DEPTOR protein, and this gel mobility was restored when nocodazole was removed from the media (Fig. 4B). This suggests that DEPTOR might be regulated by phosphorylation during completion of the cell cycle. Because DEPTOR phosphorylation is known to regulate its ability to repress mTOR signaling (*8*), we wanted to determine we might be able to overcome this reduction in DEPTOR function during cytokinesis by inducing its overexpression. To do this, we measured histone phosphorylation, which regulates and provides a convenient marker of mitotic progression (*23*). While nocodazole increased the percentage of phosphorylated S10 Histone H3 (p-H3) positive control and DEPTOR overexpressing cells similarly, release from mitotic arrest was slower in DEPTOR overexpressing cells than control cells (Fig. 4C-D). In summary, this data suggests that DEPTOR is regulated both post-translational and by localization during the later stages of the cell cycle and that inactivation or hyperactivation of mTORC1 signaling both prolong the time necessary to complete cytokinesis.

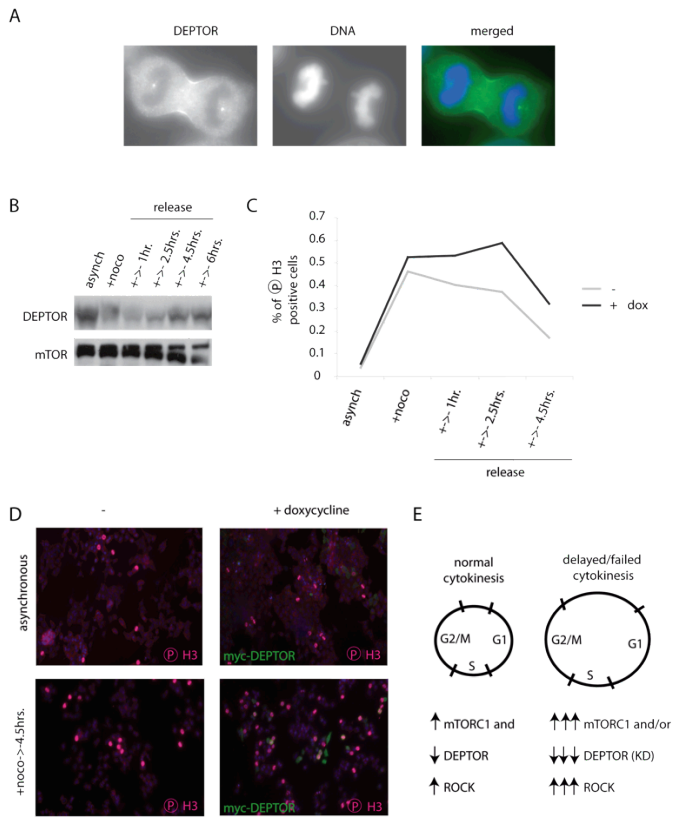

**Fig. 4. Further characterization of DEPTOR during cytokinesis.**
(A) HeLa cells were processed in an immunofluorescence assay to detect DEPTOR (green), costained with DAPI for DNA content (blue), and imaged. (B) HeLa cells were treated with 100ng/µl nocodazole or vehicle for 16 hours and released for the indicated times, cell lysates were prepared and analyzed by immunoblotting for DEPTOR or mTOR. (C) and (D) HeLa cells treated as in (B) were processed in an immunofluorescence assay to detect phosphorylated S10 Histone H3 (pink), myc-DEPTOR (green), costained with DAPI for DNA content (blue), and imaged. Three images for each condition were analyzed by image analysis software (CellProfiler) and mean measurements are shown in (C). Representative images are shown in (D). (E) Depiction of mTORC1-DEPTOR activity during the cell cycle.

## Discussion

Here we find that increased mTORC1 causes polyploidy due to defective cytokinesis that is correlated with increased Rho-ROCK-mediated phosphorylation of multiple effectors including MLC2, Mypt1, and Cofilin. While there is some evidence that mTOR regulates Rho-dependent processes in mammalian cells (*24, 25*), our work, to our knowledge, is the first to demonstrate that mTORC1 regulates cytokinesis via activation of Rho-ROCK signaling. One question still remaining is whether mTORC1 and not mTORC2 controls all mTOR-dependent, Rho signaling processes. Recently, it has been argued that, on the contrary, many of the effects of mTOR on the actin cytoskeleton are due to regulation of TORC2 and not TORC1 signaling (*24*). While our work does not directly address these claims, there is precedence with other TOR-dependent outputs that TORC1 and TORC2 might overlap in function more than is readily appreciated. In yeast, Sch9 is a target of TORC1 and controls longevity in a manner analogous to one of its homologs in higher organisms, Akt, which is an mTORC2, and not mTORC1, substrate (*26-28*). Also, loss-of-function alleles of LST8, a component of both TORC1 and TORC2, regulates rapamycin-sensitive, GAP1 permease sorting in yeast, but in mammals, does not only significant impair mTORC1 function and rather is essential for mTORC2-Akt activity (*29, 30*).

Future studies will undoubtedly be needed to better understand the mechanisms by which TOR regulates Rho function. Recently, it has been shown that rapamycin downregulates the translation of ROCK1(*31*). Whereas, our data suggests that mTORC1 also regulates the Rho pathway upstream of ROCK (Fig. 3E). Because TOR is known to regulate numerous effectors by their phosphorylation (*32*), one interesting possibility stemming from our work is to test whether mTORC1 regulates the phosphorylation state of a Rho-GAP or Rho-GEF. By answering this as well as other questions of how TOR and Rho work, it is exciting to consider that understanding of one of biology's central questions 'how cell size is coordinated with cell division' will be closer in reach.

Experiments in Figure 1 were performed by T.R.P.
Experiments in Figure 2 were performed by T.R.P. and assisted by E.V.
Experiments in Figure 3 were performed by T.R.P.
Experiments in Figure 4 were performed by T.R.P.

**Experimental procedures**

**Materials**
Reagents were obtained from the following sources: Rho Assay Reagent (Rhotekin RBD agarose), and antibodies to DEPTOR, raptor, Rho, phospho-T696 Mypt1 from Millipore; mouse monoclonal DEPTOR antibody from Novus Biologicals; antibodies to phospho-S10 Histone H3, mTOR, actin, as well as HRP-labeled secondary antibodies from Santa Cruz Biotechnology; antibodies to mTOR, phospho-S19 MLC2, phospho-S3 Cofilin, Cofilin, and the c-MYC epitope from Cell Signaling Technology; α-tubulin antibodies, Y-27632, and nocodazole from Sigma Aldrich; DMEM from SAFC Biosciences; rapamycin from LC Labs; PreScission protease from Amersham Biosciences; pTREQ Tet-On vector from Clontech; FuGENE 6 and Complete Protease Cocktail from Roche; Alexa Fluor 488 and 568 Phalloidin, Propidium Iodide, and inactivated fetal calf serum (IFS) from Invitrogen. Torin1 was kindly provided by Nathaniel Gray (Harvard medical School).

**Cell Lines and Cell Culture**
HeLa, HEK-293T, and MEFs were cultured in DMEM with 10% Inactivated Fetal Bovine Serum (IFS). p53-/- MEFs were kindly provided by David Kwitakowski (Harvard Medical School). The HeLa cell line with doxycycline-inducible DEPTOR expression was generated by retroviral transduction of HeLa that were previously modified to express rtTA with an inducible DEPTOR cDNA (8).

**cDNA Manipulations, Mutagenesis, and Sequence Alignments**
The cDNAs for DEPTOR and metap2 was previously described. The cDNA for full-length wild-type ROCK2 was kindly provided by K. Kaibuchi (33).

**Mammalian Lentiviral shRNAs**
Lentiviral shRNAs to the indicated human and mouse genes were previously described (8, 28).

shRNA-encoding plasmids were co-transfected with the Delta VPR envelope and CMV VSV-G packaging plasmids into actively growing HEK-293T using FuGENE 6 transfection reagent as previously described (28). Virus containing supernatants were collected at 48 hours after transfection, filtered to eliminate cells, and target cells infected in the presence of 8 µg/ml polybrene. For all cell types, 24 hours after infection, the cells were split into fresh media, selected with 1 µg/ml puromycin. Five days post-infection, shRNA-expressing cells were analyzed or split again and analyzed 2-3 days later. All shRNA-expressing cells were analyzed at 50-75% confluence.

**Cell Lysis and Immunoprecipitations**
All cells were rinsed with ice-cold PBS before lysis. All cells were lysed with Triton-X 100 containing lysis buffer (40 mM HEPES [pH 7.4], 2 mM EDTA, 10 mM sodium pyrophosphate, 10 mM sodium glycerophosphate, 150 mM NaCl, 50 mM NaF, 1% Triton-X 100, and one tablet of EDTA-free protease inhibitors [Roche] per 25 ml). The soluble fractions of cell lysates were isolated by centrifugation at 13,000 rpm for 10 min in a microcentrifuge. For measurement of the Rho-GTP loading state, Rho Assay Reagent was used according to the manufacturer's protocol (Millipore). To observe gel mobility shifting in DEPTOR, 8% Tris Glycine gels (Invitrogen) were used. For all other applications, 4-12% Bis-Tris gels (Invitrogen) were used.

**cDNA Transfection**
To examine the effects of ROCK2 overexpression on endogenous S19 MLC2 phosphorylation, 500,000 HEK-293T were plated in 6 cm culture dishes in DMEM/10% IFS. 24 hours later, cells were transfected with 1 µg of the indicated pRK5-based cDNA expression plasmids. All cells were lysed at 50-75% confluence 24 hours after transfection.

## Immunofluorescence Assays

25,000-100,000 cells were plated on fibronectin-coated glass coverslips in 12-well tissue culture plates, rinsed with PBS once and fixed for 15 minutes with 4% paraformaldehyde in PBS warmed to 37°C. The coverslips were rinsed three times with PBS and permeabilized with 0.2% Triton X-100 in PBS for 15 minutes. After rinsing three times with PBS, the coverslips were blocked for one hour in blocking buffer (0.25% BSA in PBS), incubated with primary antibody in blocking buffer overnight at 4°C, rinsed twice with blocking buffer, and incubated with secondary antibodies (diluted in blocking buffer 1:1000) and/or Phalloidin for one hour at room temperature in the dark. The coverslips were then rinsed twice more in blocking buffer and twice in PBS, mounted on glass slides using Vectashield containing DAPI (Vector Laboratories), and imaged with a 10x or 63X objective using epifluorescence microscopy. Quantification of the percentage of p-H3 positive cells was performed with CellProfiler (www.cellprofiler.org) using 10x images. After illumination correction, the nuclei were automatically identified using the DAPI staining. P-H3 positive cells were defined by threshold intensity. The percentage of p-H3 positive cells was then determined by the quotient of the total number of nuclei above that threshold intensity divided by total number of nuclei in the field.

## Live Cell Imaging

HeLa cells were plated on glass-bottom dishes (MatTek). Cells were imaged with an ORCA-ER camera (Hammamatsu) attached to a Nikon TE2000 microscope. All images were collected, measured, and compiled with the aid of Metamorph imaging software (Molecular Devices) and Adobe Photoshop. For time-lapse imaging, cells were kept at 37°C with the aid of a Solent incubation chamber (Solent Inc.).

## Fluorescence Activated Cell Sorting

Exponentially growing cells were treated as indicated, trypsinized, washed once in ice-cold PBS and fixed overnight in 1 ml 70% ethanol on ice. Cells were washed once in PBS and incubated in PBS containing 70 µM propidium iodide and 10 mg/ml RNAse A at 37°C for 30 minutes. Flow cytometry analysis was performed using a Becton-Dickinson FACScan machine and CellQUEST DNA Acquisition software.

## Acknowledgements

The authors thank Glenn Paradis for technical assistance. We thank members of the Sabatini lab for helpful discussions. This work was supported by fellowships from the American Diabetes Association and Ludwig Cancer Fund to T.R.P.; D.M.S. is an investigator of the Howard Hughes Medical Institute and is additionally supported by grants from the National Institutes of Health and awards from the Keck Foundation and LAM Foundation; None of the authors have a conflict of interest related to the work reported in this manuscript.